\documentclass[aps,prl,twocolumn]{revtex4-1}
\usepackage[bookmarks,colorlinks,citecolor=blue,linkcolor=blue]{hyperref}

\usepackage{amssymb}
\usepackage{amsmath,bm}
\usepackage{graphicx,color}
\usepackage{bbm}
\usepackage[bottom]{footmisc}
\usepackage{multirow}

\begin{document}
\title{Absorbing-state transitions in particulate systems under spatially varying driving}
\author{Bhanu Prasad Bhowmik}
\email{bhowmikbhanuprasad592@gmail.com}
\author{Christopher Ness}
\affiliation{School of Engineering, University of Edinburgh, Edinburgh EH9 3JL, United Kingdom}

\begin{abstract}
We study the absorbing state transition in particulate systems under spatially inhomogeneous driving
using a modified random organization model.
For smoothly varying driving the steady state results map onto
the homogeneous absorbing state phase diagram, with the position of the boundary between absorbing and diffusive states being insensitive to the driving wavelength.
Here the phenomenology is well-described by a one-dimensional continuum model that we pose.
For discontinuously varying driving the position of the absorbing phase boundary and the exponent characterising the fraction of active particles are altered relative to the homogeneous case.  
\end{abstract}

\maketitle

\paragraph{Introduction.}
The response of soft materials to cyclic deformation is a subject of intense research due to its relevance
to liquefaction~\cite{huang2013review}, hopper unblocking~\cite{janda2009unjamming} and yielding~\cite{bonn2017yield}.
This process gives rise to phenomena of fundamental interest,
including absorbing state transitions and self-organized criticality~\cite{Pine2005, cortePineNature2008, CortePRL2009, MansaKandulaPREA2014, FioccoFoffiSastryPRE2013},
mechanical annealing~\cite{HimangshuPNAS2021}
and
fatigue failure~\cite{BHPEPL2022}.
Most work to date focuses on homogeneous cases in which the externally applied driving shear rate is spatially uniform,
though in practical scenarios the driving is almost certain to be inhomogeneous.
Recent studies in various soft materials~\cite{TigheNonLocalPRL, bhowmik2023scaling, BocquetNonLocalNature, KamrinKovalPRL2012, PinakiAndJuergenPRE2014} reveal that the rheology of systems under homogeneous and inhomogeneous driving often differs qualitatively, so that the response to the latter cannot necessarily be fully predicted based on the former.
\begin{figure*}
\includegraphics[trim = 0mm 35mm 0mm 0mm, clip,width=0.985\textwidth,page=1]{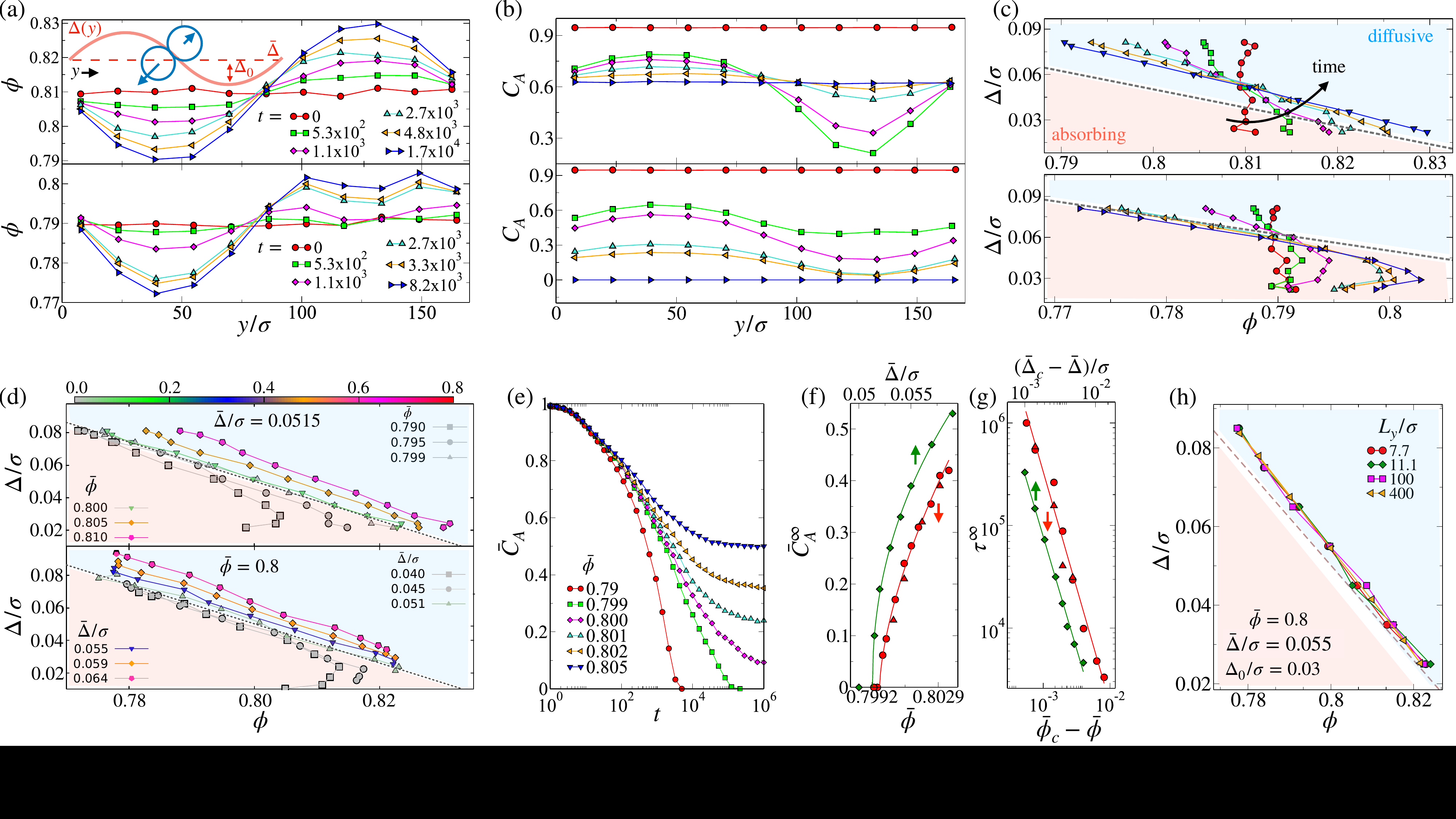}
\caption{
Absorbing-state transition in inhomogeneously driven particulate material.
Shown are the time evolution of (a) area fraction $\phi(y)$ and (b) active particle concentration $C_{A}(y)$, for systems with steady states lying above (top, $\bar{\phi}=0.81$) and below (bottom, $\bar{\phi}=0.79$) the absorbing phase boundary, with $\bar{\Delta}/\sigma = 0.0515$, $\Delta_0/\sigma = 0.03$ and $L_y/\sigma=160$.
The Inset of (a) is a schematic of the dynamics;
(c) The $\Delta - \phi$ phase space showing parametrically $\phi(y)$ and $\Delta(y)$ at progressing time.
The dashed line represents the homogeneous phase boundary separating absorbing (pink) and diffusive (blue) states
(legend in (a) applies also to (b) and (c));
(d) Steady state $\Delta - \phi$ lines for {$\bar{\Delta}=0.0515$} at a range of $\bar{\phi}$ (top) and {$\bar{\phi}=0.8$} at a range of $\bar{\Delta}$ (bottom), with $\Delta_0=0.03$.
Colorbar represents the magnitude of $C_{_A}(y)$ (which is approximately uniform in space);
(e) Time evolution of mean active particle concentration $\bar{C}_{A}$ for various $\bar{\phi}$ keeping $\bar{\Delta}$, $\Delta_0$ fixed;
(f) Variation of $\bar{C}^\infty_{A}$ with $\bar{\phi}$ at fixed $\bar{\Delta}$ (red) and  $\bar{\Delta}$ at fixed $\bar{\phi}$ (green (circles are $L_y/\sigma=120$; triangles are $L_y/\sigma=11$));
(g) Variation of the time $\tau^{\infty}$ required to reach absorbing states when below the phase boundary, with $\bar{\phi}$ (green) and  $\bar{\Delta}$ (red).
Solid lines in (f) and (g) are fits to $f(x) = f_0\left(x-x_c\right)^a$.
(h) Example steady state $\Delta-\phi$ lines (parameters in legend) showing independence of $L_y$ in the diffusive regime.
} 
\label{fig1}
\end{figure*}

Here we focus on the non-equilibrium absorbing state transition in athermal particulate systems below jamming.
Under cyclic deformation
with driving amplitude $\gamma$ less than a volume (or area) fraction $\phi$ dependent threshold $\gamma_c(\phi)$,
such systems attain special configurations which reappear precisely after full cycles of deformation,
so that,
when viewed stroboscopically (with a period of 1 or occasionally more cycles~\cite{schreck2013particle}), the system does not explore configuration space.
Above $\gamma_c$ there exist active particles $(C_{A})$ that do not return to their initial positions so that the self-diffusion coefficient is nonzero.
These states are separated by a critical line on the $\phi - \gamma$ phase diagram,
exhibiting an equilibrium-like continuous transition with $C_{A}$ serving as the order parameter.

To study the physics of this transition it has proven useful to explore  `random organization' models \cite{cortePineNature2008, TjhungAndBethierPRL2015, GCBPRL2023,HexnerPRL2015,ChrisNessPRL2020AbsState},
whose predictions
suggest it belongs to the conserved directed percolation (CDP) class,
though this is altered in the presence of multiple or mediated interactions \cite{ShankarGhoshPRL2022,MariPRE2022}.
Thus far, such models are reported for homogeneous driving only,
yet in practical scenarios involving cyclic driving of particulates
the driving rate may inherently be non-uniform.
More broadly the role of spatial inhomogeneity in processes with absorbing states is relevant to regionally varying immunity levels in disease spreading scenarios,
or to reaction-diffusion problems with spatially varying catalytic activity.
In order to gain fundamental insight into this more complex class of problems it is thus crucial to determine whether the presence of spatially varying driving forces affects the nature and position of absorbing state transitions in discrete systems,
and, importantly, whether homogeneous measurements predict the properties of inhomogeneous systems.

We use a modified isotropic random organization model~\cite{MilzSchmiedebergPRE2013,TjhungAndBethierPRL2015} to study the absorbing state transition under inhomogeneous driving.
The control parameters are the overall area fraction $\bar{\phi}$ of particles of diameter $\sigma$ and their per-step displacement $\Delta$.
The latter is an analogue of the oscillatory strain amplitude $\gamma$,
and we impose spatial dependence $\Delta(y)$ to represent \emph{e.g.} the spatially-varying strain rate present in a flowing system.
For diffusive states we observe in the results a spatial inhomogeneity with the local area fraction at $y$, $\phi(y)$, being lower in regions where the driving amplitude $\Delta(y)$ is large.
For smoothly varying driving ($\partial \Delta/\partial y$ exists at every point)
with wavelength $L_y/\sigma=7.7-400$, the local active particle concentration $C_{A}(y)$
matches that obtained under homogeneous conditions at the same $\Delta(y)$ and $\phi(y)$,
indicating that the presence of gradients does not influence the properties of the $\Delta-\phi$ phase diagram.
We thus explore a continuum model modified to account for inhomogeneous driving,
and find it supports our simulation results,
albeit with the expected mean-field exponents \cite{SLuebeckScaling2004,SriramGautomMenonPRE}.
Conversely, for discontinuous $\Delta(y)$
inhomogeneous effects play a role and the properties of the phase diagram depend on the length $L_c$ over which the driving remains uniform.
For $L_c/\sigma \lesssim 20$,
the position of the absorbing boundary depends on $L_c$,
while for the extreme case in which $\Delta = 0$ for a small fraction of particles $C_p$, the position and exponent of the transition
are $C_p$-dependent, indicating a change to the universality class.


\paragraph{Simulation details.}
We simulate $N=5000-30000$ disks with diameter chosen from a Gaussian with mean $\sigma$ and standard deviation $0.2\sigma$ in a box of area $L_x \times L$
(with $L$ an integer multiple of $L_y$, defined below).
Initially random configurations having widespread particle overlaps evolve according to the following deterministic rules.
Particles not overlapping with any other are inactive and do not move.
Particles with $Z>0$ overlapping neighbors are active and their positions $\bm{r}_i$ are updated following
$\bm{r}_i\left(t+\delta t\right) = \bm{r}_i\left( t\right) + \Delta_i \delta t \sum_{j=1}^{Z} \bm{n}_{ij}$, where $\bm{n}_{ij}$ are unit vectors pointing from particle $i$ to each of its contacts $j$, $\Delta_i$ is the kick size for the $i^{th}$ particle, and $\delta t = 1$.
The overall area fraction is $\bar{\phi}=\frac{\sum_{i=0}^{N} \pi \sigma_i^2}{4L_xL}$.
We produced the $\Delta-\phi$ phase diagram for spatially-invariant $\Delta$ (finding results consistent with~\cite{TjhungAndBethierPRL2015,pham2016origin})
and refer to it in what follows.
For inhomogeneous driving we initially let $\Delta_i$
be smoothly varying as $\Delta_i(y_i) = \bar{\Delta} + \Delta_0\sin(\frac{2\pi y_i}{L_y})$,
Fig.~\ref{fig1}(a) [Inset].
The inputs to our model are thus $\bar{\phi}$, $\bar\Delta$, $\Delta_0$ and $L_y$,
while the observables we compute are the spatial profiles of the area fraction $\phi(y)= \frac{1}{L_x}\int_0^{L_x}\phi(x,y)dx$
and the active particle concentration $C_{A}(y) = N_A(y)/N(y)$,
with $N_A(y)$ and $N(y)$ the number of active and total number of particles at position $y$.
Coarse grained profiles are computed by binning particles in $y$
before averaging the particle properties in each bin.
Results are averaged across 50 realisations for each parameter set.
We define
the order parameter
$\bar{C}_A = \frac{1}{L}\int_0^{L}C_A(y)dy$,
and $\bar{C}_A^{\infty}$ as its steady state value.

\paragraph{Smoothly varying $\Delta(y)$.}
Using sinusoidal $\Delta(y)$ we explore a range of $\bar{\Delta}$ (letting $\Delta_0 = 0.03\sigma$), $\bar{\phi}$ and $L_y$,
to investigate conditions that in principle span across the homogeneous absorbing state phase boundary.
Model predictions are shown in Fig.~\ref{fig1}.
Inhomogeneous driving produces steady states with spatially varying $\phi$,
with regions of higher $\Delta(y)$ having lower $\phi(y)$, Fig.~\ref{fig1}(a),
in agreement with experiment~\cite{Migration2,Migration1} and simulations with more detailed physics~\cite{TigheNonLocalPRL,bhowmik2023scaling,GillissenNessPRL2020}.
The local active particle concentration $C_{A}(y)$ varies in the opposite direction to $\phi(y)$ at small $t$,
before reaching steady states that are spatially uniform, Fig.~\ref{fig1}(b).
For ($\bar{\phi}$, $\bar{\Delta}$) above the homogeneous phase boundary (top panels of (a),(b)),
the inhomogeneous system remains diffusive at long times with $C_{A}(y) > 0$,
whereas below the homogeneous phase boundary $C_{A}(y) = 0$ everywhere in the system.
The system never produces mixed absorbing-diffusive steady states with $C_{A}(y)$ only locally vanishing.

Each inhomogeneous simulation produces 
a set of parametric points corresponding to lines across the $\Delta-\phi$ phase diagram,
bounded by $\Delta = \bar{\Delta} \pm \Delta_0$, Fig.~\ref{fig1}(c). 
Initial configurations with uniform $\phi(y)$
appear as vertical $\Delta-\phi$ lines,
before evolving with time.
Since the local activity at $y$ is controlled by both $\phi(y)$ and $\Delta(y)$,
particles in regions with higher $\Delta$ will have higher mobility and move to regions with lower $\Delta$, resulting in an increment of the area fraction in those regions.
This consequently increases the activity of those regions.
Thus, regions with initial local parameters below the absorbing phase boundary can increase in $\phi(y)$ and thus become active.
This process manifests as a counterclockwise rotation of $\Delta-\phi$, Fig.~\ref{fig1}(c) (top).
For ($\bar{\phi}, \bar{\Delta}$) above the homogeneous phase boundary, after this transient the system attains a diffuse steady state with $\Delta(y)$ and $\phi(y)$ balanced such that $C_A(y)$ becomes spatially uniform.
For ($\bar{\phi}, \bar{\Delta}$) below the phase boundary,
the large-$\Delta$ regions
do not inject enough particles to the small-$\Delta$ regions to trigger activity,
thus leaving those areas with the same conditions as at the beginning of the simulation and the resultant hook shape in Fig.~\ref{fig1}(c) (bottom). 

Figure~\ref{fig1}(d) shows steady state $\Delta-\phi$ lines as functions of $\bar{\phi}$ (top) and $\bar{\Delta}$ (bottom).
The colorbar stands for the magnitude of ${C}_A(y)$,
which decreases as $\left(\bar{\phi}, \bar{\Delta}\right)$ approaches the phase boundary from within the diffusive phase,
and vanishes in the absorbing state.
Interestingly,
the location of the critical line at which $C_A(y)$ vanishes overlaps with the homogeneous absorbing phase boundary (dashed lines, Fig.~\ref{fig1}(d)),
so that in principle a single inhomogeneous simulation can identify the entire homogeneous phase boundary.
In Fig.~\ref{fig1}(e) the variation of $\bar{C}_{A}$ with time is shown for various $\bar{\phi}$ at fixed $\bar{\Delta}$, $\Delta_0$.
As in the homogeneous case,
$\bar{C}_{A}$ vanishes at low $\bar{\phi}$,
indicating that absorbing states exist,
whereas above it saturates to non-zero steady states $\bar{C}_A^\infty$ which we characterise by fitting the data to $C_{_A}(t) = c_0e^{-(t/t_0)^{c_1}} + \bar{C}_{_A}^\infty$, where $c_0, t_0$  and $c_1$ are fitting parameters. 

As $\bar{C}_A^\infty$ serves as our order parameter,
we investigate the details of the absorbing state transition by examining its dependence on the parameters.
The measured values follow $\bar{C}_A^\infty \sim \left(\bar{\phi} - \bar{\phi}_c\right)^{\beta\phi}$ for fixed $\bar{\Delta}$ and $\bar{C}_A^\infty \sim \left(\bar{\Delta} - \bar{\Delta}c\right)^{\beta_{\Delta}}$ for fixed $\bar{\phi}$, Fig.~\ref{fig1}(f),
and we find $\beta_{\phi} = 0.64$, in agreement with Refs \cite{SLuebeckScaling2004,GCBPRL2023},
$\bar{\phi}_c =  0.7996$,
and $\beta_{\Delta} = 0.45$, close to the values reported by Ref~\cite{cortePineNature2008}, with $\bar{\Delta}_c = 0.0513$.
Additionally, as the system approaches the critical line from the absorbing phase, the time required to reach steady state diverges as $\tau^\infty \sim \left(\bar{\phi} - \bar{\phi}_c\right)^{\nu_{\phi}}$ for fixed $\bar{\Delta}$ and $\tau^\infty \sim \left(\bar{\Delta} - \bar{\Delta}c\right)^{\nu_{_\Delta}}$ for fixed $\bar{\phi}$,~Fig.~\ref{fig1}(g).
We find $\nu_{\phi} = 1.37$ with $\bar{\phi}_c = 0.7994$,
and $\nu_{\Delta} = 1.33$ with $\bar{\Delta}_c = 0.0513$,
in close agreement with previous studies~\cite{SLuebeckScaling2004,GCBPRL2023,cortePineNature2008}.
Thus the position and nature of the phase boundary obtained under inhomogeneous conditions matches the homogeneous case. We verified that this holds for $L_y/\sigma=7.7-400$, Figs.~\ref{fig1}(f)-(h).

\begin{figure}
\includegraphics[trim = 0mm 0mm 260mm 0mm, clip,width=0.485\textwidth,page=2]{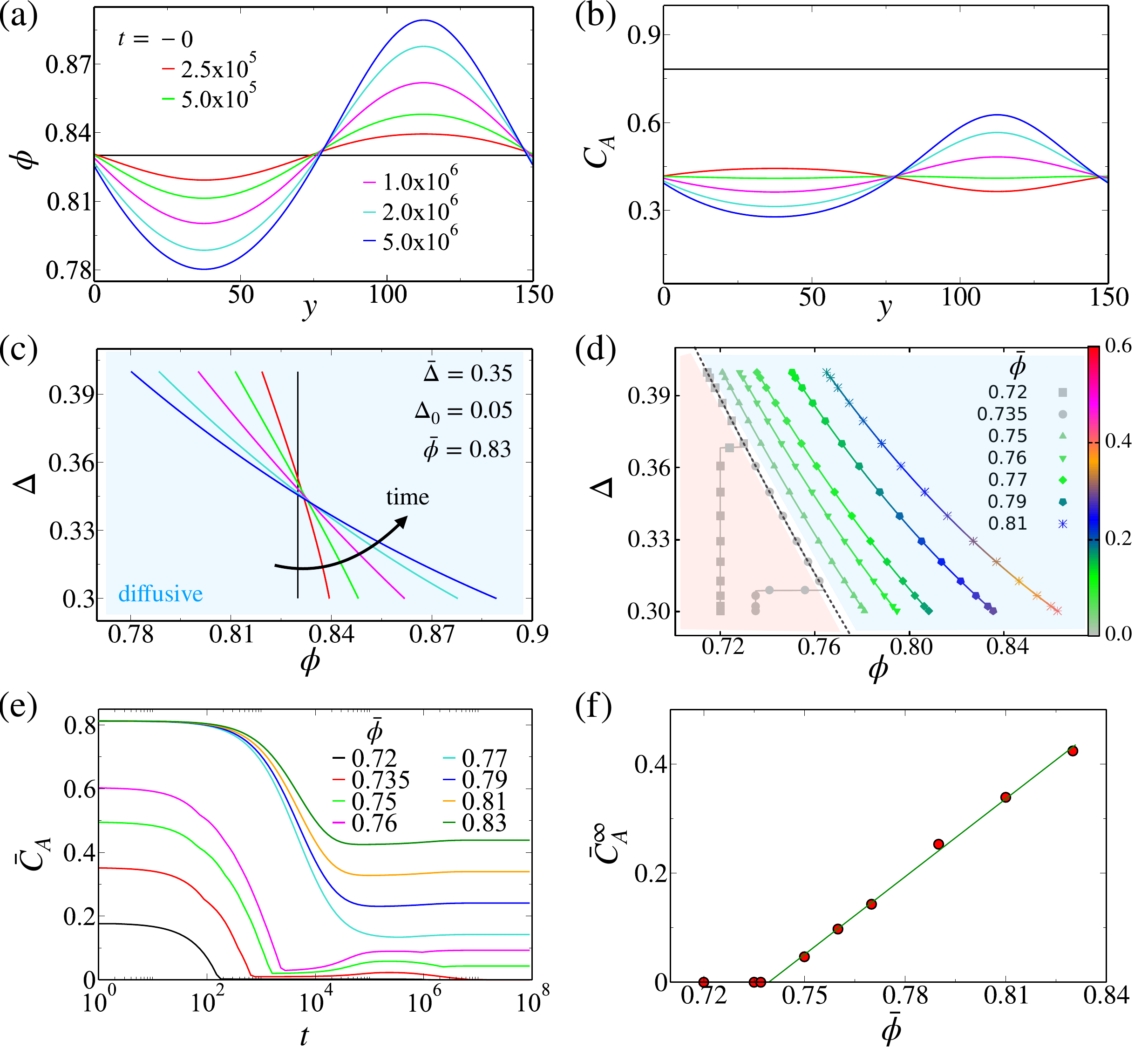}
\caption{Predictions of continuum model.
Shown are the time evolution of (a) $\phi(y)$,
(b) $C_{A}(y)$
and (c) the $\Delta - \phi$ line,
from initialization up to a steady state in the diffusive regime (parameters in Legend of (c)).
The color scale in (a) applies also to (b) and (c);
(d) Steady state $\Delta - \phi$ lines for various $\bar{\phi}$ at fixed profiles of ${\Delta}(y)$;
colorbar shows the magnitude of $C_A(y)$.
(e) Time evolution of $\bar{C}_{A}$ for a range of $\bar{\phi}$ spanning the absorbing phase boundary;
(f) Variation of $\bar{C}^\infty_{A}(y)$ with $\bar{\phi}$ keeping the ${\Delta}(y)$ profile fixed, with solid line as in Fig.~\ref{fig1}(f).}
\label{fig2}
\end{figure}

\paragraph{Continuum model.}
Given that the phase boundary is governed by local conditions only,
we introduce a continuum model (Manna class \cite{VespignaniZapperiPRL1997, ZapperiPRL1998} without noise \cite{Hipke2009}) modified for inhomogeneous driving by introducing convection in the system.
Herein $\phi_{_A}(y,t)$ and $\phi_{_I}(y,t)$ are the area fraction of active and inactive particles, related through $\phi(y,t) = \phi_{_A}(y,t) + \phi_{_I}(y,t)$, $\dot{\phi} = \dot{\phi}_{_A} + \dot{\phi}_{_I}$,
and with $C_A = \phi_{A}/\phi$. The dynamics are given by
\begin{equation}
\begin{aligned}
\dot{\phi}_{_A} ={} & \nabla\left( \mathcal{D}\nabla \phi_{_A} \right) + \nabla \left(\nabla \Delta \phi_{_A}\right)  \\
      & + \alpha \phi_{_A} (\phi - \phi_{_A}) - \beta \phi_{_A}\left(1-\phi\right)\text{,}
\end{aligned}
\label{EqC_A}
\end{equation}
\begin{equation}
\dot{\phi}_{_I} = - \alpha \phi_{_A} (\phi - \phi_{_A}) +  \beta \phi_{_A}\left(1-\phi\right) \text{,}
\label{EqC_I}
\end{equation}
where $\nabla = \frac{\partial}{\partial y}$, and $\alpha$ and $\beta$ are positive coefficients. $\alpha$ represents activation of particles by interaction with active neighbours; $\beta$ represents isolated death (we omit caging~\cite{XuSchwarzPRE2013}).
The convection term in Eq.~\ref{EqC_A} arises due to spatially-varying driving.
The diffusion coefficient is taken as $\mathcal{D}(y) \sim \Delta^2(y)/\tau$, with the time scale $\tau$ taken as unity.
The phase boundary can be estimated from the condition $\phi_{_A} = 0$, $\nabla \phi_{_A} = 0$ and $\dot{\phi}_{_A} = 0$.
The coefficient $\alpha$ is directly related to $\Delta$, and for simplicity we choose $\alpha \sim \Delta$.
Thus the critical mean kick size $\bar{\Delta}_c$ representing the phase boundary at a fixed $\bar{\phi}$ is
$\bar{\Delta}_c = \beta/\bar{\phi} - \beta$\text{.}
\begin{figure}
\includegraphics[trim = 0mm 0mm 238mm 0mm, clip,width=0.485\textwidth,page=3]{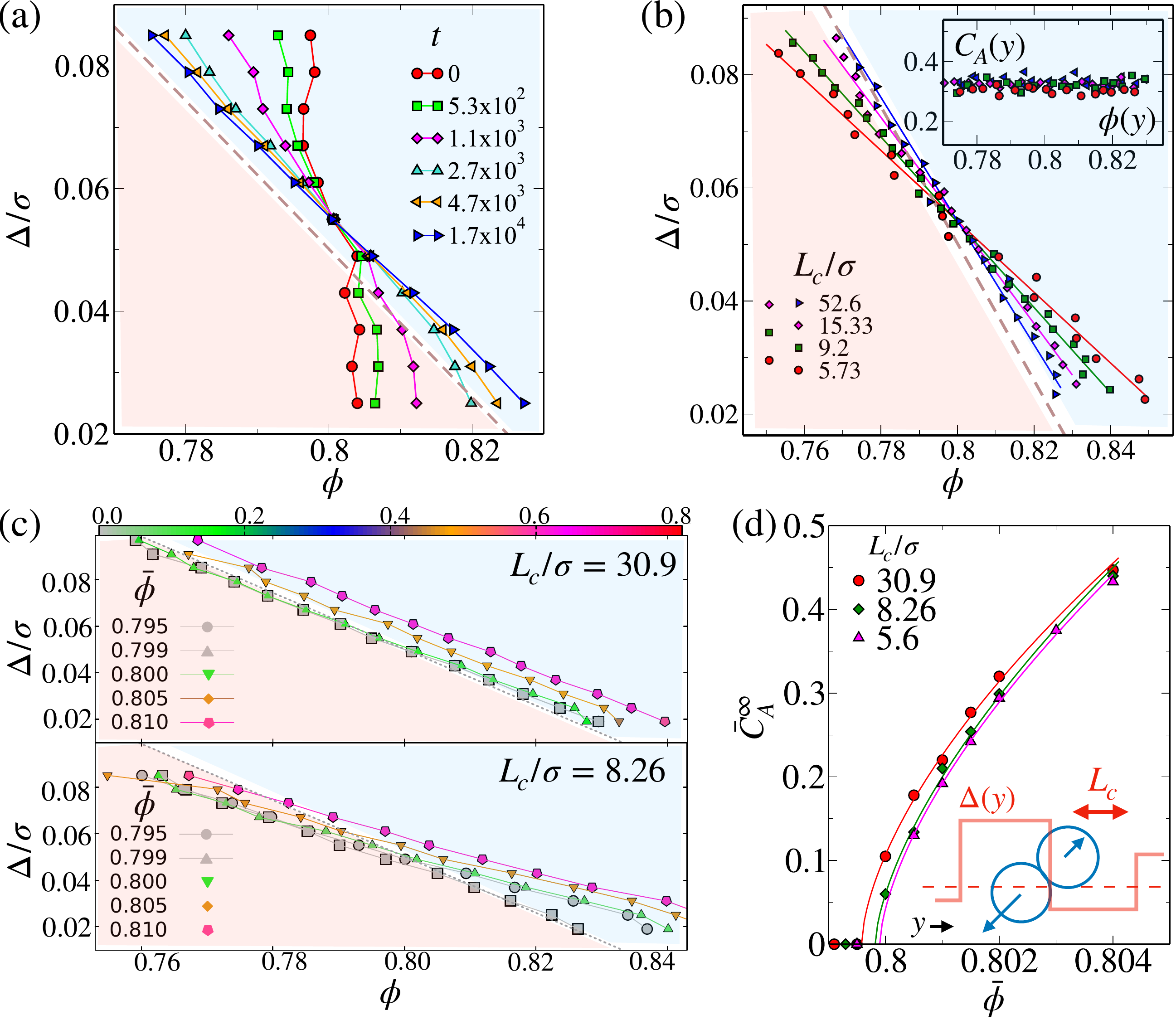}
\caption{Model predictions with discontinuous $\Delta(y)$.
(a) Time evolution of  $\Delta-\phi$ for $\bar{\phi} = 0.80$,
with
$\bar{\Delta} = 0.056$,
$\Delta_0=0.03$ and $L_c/\sigma=52.6$,
showing similar behaviour to the smoothly varying case;
(b) Steady state $\Delta-\phi$ lines for $\bar{\phi} = 0.80$, $\bar{\Delta} = 0.054$ and $\Delta_0=0.03$, as a function of cell size $L_c$.
Dashed line shows homogeneous absorbing state phase boundary.
Inset: steady state ${C}_A(y)$ as a function of $\phi(y)$ for each case verifying that they are all diffusive despite crossing the boundary;
(c) Steady state $\Delta-\phi$ lines for $\bar{\Delta} = 0.0515$,
$\Delta_0=0.03$ at different $\bar{\phi}$ for $L_c=30.9$ (top) and $L_c=8.26$ (bottom). Grey dotted lines are the homogeneous absorbing phase boundary and the colorbar represents $C_A(y)$; (d) Variation of $\bar{C}^\infty_{A}$ with $\bar{\phi}$,
keeping $\bar{\Delta} = 0.0515$, showing the changing position of the boundary with $L_c$.
Solid lines follow the form in Fig.~\ref{fig1}(f).
Inset: schematic of model dynamics.
}    
\label{fig3}
\end{figure}
We solve these coupled equations numerically,
with $\Delta(y) = \bar{\Delta} + \Delta_0\sin(\frac{2\pi y}{L_y})$ and initial conditions $\phi(y) = \bar{\phi}=0.83$, $\phi_A(y) = \phi_A^0=0.65$
using $\bar{\Delta} = 0.35$, $\Delta_0 = 0.05$.
Shown in Figs.~\ref{fig2}(a)-(c) are transients of $\phi(y)$ and $C_{A}(y)$ in the diffusive regime,
showing qualitative agreement with Figs.~\ref{fig1}(a)-(c) (top).
In Fig.~\ref{fig2}(d) are steady state $\Delta - \phi$ lines spanning the absorbing phase boundary,
qualitatively matching the random organization model predictions in Fig.~\ref{fig1}(d).
In the absorbing state (gray lines) the hook shape in $\Delta-\phi$ appears as in the simulation, though with a sharper curve and a vertical profile at lower $\Delta$ where the dynamics cease quickly.
The evolution of $\bar{C}_A$ with time is shown in Fig.~\ref{fig2}(e),
approaching 0 and finite values respectively below and above the phase boundary with the time scale $\tau$ diverging at the transition.
We find $\bar{C}_{_A}^{\infty} \sim \left(\bar{\phi} - \bar{\phi}_c\right)^{\beta_c}$,
with $\beta_c = 1$ agreeing with previously reported results~\cite{XuSchwarzPRE2013,SriramGautomMenonPRE}.


\paragraph{Discontinuous $\Delta(y)$.}
We next scrutinize the random organization model under discontinuously varying $\Delta(y)$.
To do so we divide the system into cells of area $L_c^2$,
each having $\Delta = \bar{\Delta} + \Delta_r$
(Fig.~\ref{fig3}(d) [Inset]),
with $\Delta_r$ a random number chosen uniformly in the range $\pm\Delta_0$,
ensuring $\sum^{N_c}_{i=1}\Delta_r = 0$, where $N_c$ is the number of cells.
Model predictions are shown in Fig.~\ref{fig3}.

The transients observed for exemplar model parameters match qualitatively those from the smoothly varying model,
Fig.~\ref{fig3}(a),
while steady state $\Delta-\phi$ data deviate significantly from the homogeneous absorbing state phase boundary,
Fig.~\ref{fig3}(b).
As in the smoothly-varying scenario the active particles are distributed uniformly over the system in the diffusive regime, Fig.~\ref{fig3}(b) [Inset].
For small $L_c$, states
that would be deep within the absorbing state
remain active.
This indicates that under discontinuously varying driving the absence or presence of an absorbing state cannot be determined using local knowledge of $\phi(y)$ and $\Delta(y)$: the size of the containing cell matters.
Presumably particles near cell edges experience dynamics not described by Eqs.~\ref{EqC_A}-~\ref{EqC_I} due to proximal discontinuities in $\Delta$.
Indeed as $L_c$ increases (and fewer particles are near cell edges), the data
tend towards a diffusive line parallel to the homogeneous phase boundary as would be expected
for locally-governed dynamics.
In Fig.~\ref{fig3}(c) steady state $\Delta -\phi$ lines are shown for $\bar{\Delta} = 0.0515$, $\Delta_0=0.03$ and a set of $\bar{\phi}$ crossing the homogeneous phase boundary (gray dotted line), for $L_c/\sigma = 30.9$ (top) and $L_c/\sigma = 8.26$ (bottom).
The overall trend is similar to smoothly varying $\Delta(y)$, but importantly the position of the absorbing phase boundary under randomly varying $\Delta$ does not match the homogeneous case, especially at small $L_c$ 
(and indeed the large $L_c$ and $L_y$ cases do not appear to converge).
Figure~\ref{fig3}(d) shows the variation of $\bar{C}_A^\infty$ with $\bar{\phi}$ at fixed $\bar{\Delta}$ for a range of $L_c$.
The exponents match those in Fig.~\ref{fig1}(f), but crucially but there is dependence of the critical point $\bar{\phi}_c$ on $L_c$.

\begin{figure}[b]
\includegraphics[trim = 0mm 152mm 0mm 0mm, clip,width=0.485\textwidth,page=4]{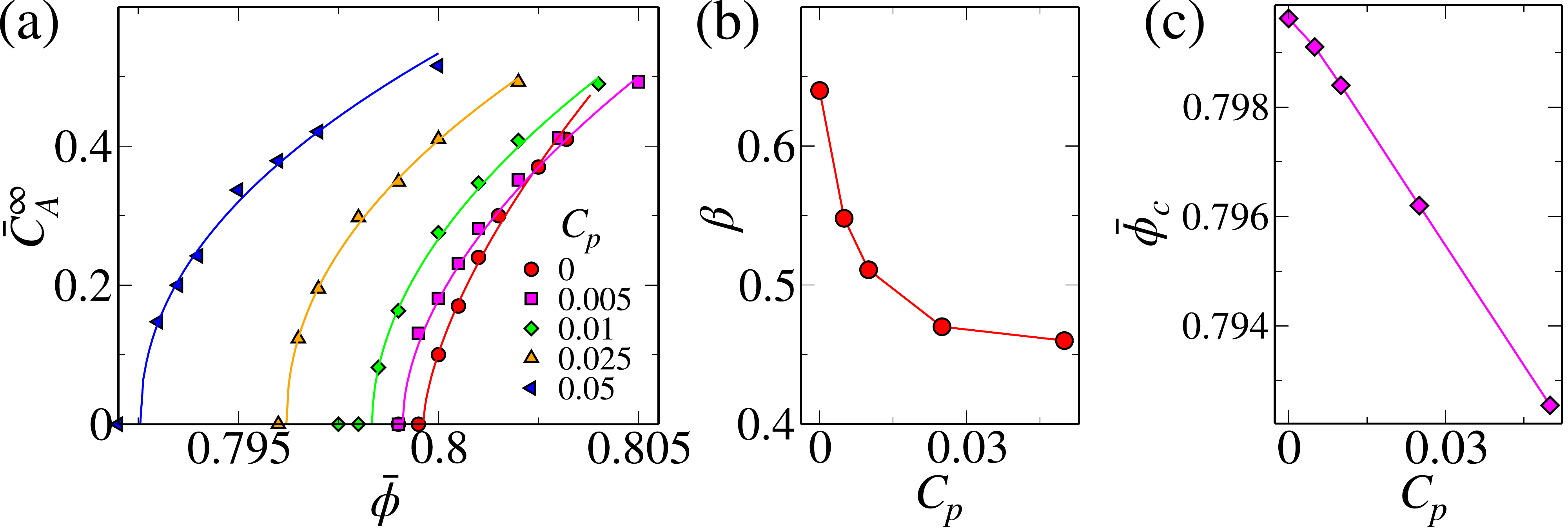}
\caption{
Absorbing state transition with permanently inactive particle fraction $C_p$.
(a) Steady state mean active particle concentration $\bar{C}^\infty_A$ as a function of $\bar{\phi}$ for varying $C_p$
with $\Delta=0.0515$;
Variation with $C_p$ of (b) the critical exponent $\beta$ and (c) the critical area fraction $\phi_c$.}     
\label{fig4}
\end{figure}
We finally test the model for an extreme case of inhomogeneity in which a small subset of particles are permanently inactive.
To do so we modify the above model by enforcing $\Delta = 0$ for a small fraction $C_p$ of particles,
chosen randomly.
The remaining particles are assigned a fixed $\Delta$,
so that the mean kick size is now given by $\bar{\Delta} = \left(1-C_p\right)\Delta$.
Particles with $\Delta = 0$ remain frozen in their initial positions, equivalent to a pinning effect used widely in fundamental studies of various systems~\cite{BiroliCammarotaPNAS2005,DBPNKPNASNexus2023,BCKPRL2019}.
We run these dynamics for a range of $C_p$ and $\bar{\phi}$,
with $\Delta=0.0515$.
Shown in Fig.~\ref{fig4}(a) is the resultant variation of the steady state $\bar{C}_A^\infty$.
The points are simulation data, and solid lines are fits to $\bar{C}^\infty_{_A} \sim \left(\phi - \phi_c\right)^{\beta}$.
The critical area fraction $\bar{\phi}_c$ and exponent $\beta$ both decrease with increasing $C_p$, Fig.~\ref{fig4}(b)-(c),
suggesting a change in the nature of criticality due to particle pinning~\cite{BKPRPRE2019}.

\paragraph{Conclusions.}
We study the absorbing state transition in particulate systems under inhomogeneous driving
using a modified random organization model.
The resulting spatial variation in area fraction
means each simulation produces a set of points in the $\Delta - \phi$ phase diagram,
which are, in steady states, either entirely below or above the absorbing phase boundary.
This behavior is reproduced with a simple continuum description of the system
incorporating a convective term to account for spatially varying driving.
When the driving varies smoothly in space the critical line separating absorbing from diffusive states is the same as for the homogeneous random organization model.
Meanwhile for discontinuously varying driving the position of the absorbing phase boundary and its properties
can deviate from the homogeneous case,
suggesting that one might tune the position of an absorbing state boundary by manipulating the hetergeneity of the (discontinuous) driving.
Our results show that the absorbing state boundary obtained under spatially homogeneous conditions can be used to predict the
features of inhomogeneous systems only when the latter involves smoothly varying driving.
Understanding the rich physics of absorbing states in particulate systems with discontinuously varying driving is thus a promising avenue of future fundamental research.

\begin{acknowledgments}
B.P.B. acknowledges support from the Leverhulme Trust under Research Project Grant RPG-2022-095;
C.N. acknowledges support from the Royal Academy of Engineering under the Research Fellowship scheme.
\end{acknowledgments}

\bibliography{ALL}

\begin{thebibliography}{37}%
\makeatletter
\providecommand \@ifxundefined [1]{%
 \@ifx{#1\undefined}
}%
\providecommand \@ifnum [1]{%
 \ifnum #1\expandafter \@firstoftwo
 \else \expandafter \@secondoftwo
 \fi
}%
\providecommand \@ifx [1]{%
 \ifx #1\expandafter \@firstoftwo
 \else \expandafter \@secondoftwo
 \fi
}%
\providecommand \natexlab [1]{#1}%
\providecommand \enquote  [1]{``#1''}%
\providecommand \bibnamefont  [1]{#1}%
\providecommand \bibfnamefont [1]{#1}%
\providecommand \citenamefont [1]{#1}%
\providecommand \href@noop [0]{\@secondoftwo}%
\providecommand \href [0]{\begingroup \@sanitize@url \@href}%
\providecommand \@href[1]{\@@startlink{#1}\@@href}%
\providecommand \@@href[1]{\endgroup#1\@@endlink}%
\providecommand \@sanitize@url [0]{\catcode `\\12\catcode `\$12\catcode
  `\&12\catcode `\#12\catcode `\^12\catcode `\_12\catcode `\%12\relax}%
\providecommand \@@startlink[1]{}%
\providecommand \@@endlink[0]{}%
\providecommand \url  [0]{\begingroup\@sanitize@url \@url }%
\providecommand \@url [1]{\endgroup\@href {#1}{\urlprefix }}%
\providecommand \urlprefix  [0]{URL }%
\providecommand \Eprint [0]{\href }%
\providecommand \doibase [0]{http://dx.doi.org/}%
\providecommand \selectlanguage [0]{\@gobble}%
\providecommand \bibinfo  [0]{\@secondoftwo}%
\providecommand \bibfield  [0]{\@secondoftwo}%
\providecommand \translation [1]{[#1]}%
\providecommand \BibitemOpen [0]{}%
\providecommand \bibitemStop [0]{}%
\providecommand \bibitemNoStop [0]{.\EOS\space}%
\providecommand \EOS [0]{\spacefactor3000\relax}%
\providecommand \BibitemShut  [1]{\csname bibitem#1\endcsname}%
\let\auto@bib@innerbib\@empty
\bibitem [{\citenamefont {Huang}\ and\ \citenamefont
  {Yu}(2013)}]{huang2013review}%
  \BibitemOpen
  \bibfield  {author} {\bibinfo {author} {\bibfnamefont {Y.}~\bibnamefont
  {Huang}}\ and\ \bibinfo {author} {\bibfnamefont {M.}~\bibnamefont {Yu}},\
  }\href@noop {} {\bibfield  {journal} {\bibinfo  {journal} {Natural hazards}\
  }\textbf {\bibinfo {volume} {65}},\ \bibinfo {pages} {2375} (\bibinfo {year}
  {2013})}\BibitemShut {NoStop}%
\bibitem [{\citenamefont {Janda}\ \emph {et~al.}(2009)\citenamefont {Janda},
  \citenamefont {Maza}, \citenamefont {Garcimart{\'\i}n}, \citenamefont {Kolb},
  \citenamefont {Lanuza},\ and\ \citenamefont
  {Cl{\'e}ment}}]{janda2009unjamming}%
  \BibitemOpen
  \bibfield  {author} {\bibinfo {author} {\bibfnamefont {A.}~\bibnamefont
  {Janda}}, \bibinfo {author} {\bibfnamefont {D.}~\bibnamefont {Maza}},
  \bibinfo {author} {\bibfnamefont {A.}~\bibnamefont {Garcimart{\'\i}n}},
  \bibinfo {author} {\bibfnamefont {E.}~\bibnamefont {Kolb}}, \bibinfo {author}
  {\bibfnamefont {J.}~\bibnamefont {Lanuza}}, \ and\ \bibinfo {author}
  {\bibfnamefont {E.}~\bibnamefont {Cl{\'e}ment}},\ }\href@noop {} {\bibfield
  {journal} {\bibinfo  {journal} {Europhysics Letters}\ }\textbf {\bibinfo
  {volume} {87}},\ \bibinfo {pages} {24002} (\bibinfo {year}
  {2009})}\BibitemShut {NoStop}%
\bibitem [{\citenamefont {Bonn}\ \emph {et~al.}(2017)\citenamefont {Bonn},
  \citenamefont {Denn}, \citenamefont {Berthier}, \citenamefont {Divoux},\ and\
  \citenamefont {Manneville}}]{bonn2017yield}%
  \BibitemOpen
  \bibfield  {author} {\bibinfo {author} {\bibfnamefont {D.}~\bibnamefont
  {Bonn}}, \bibinfo {author} {\bibfnamefont {M.~M.}\ \bibnamefont {Denn}},
  \bibinfo {author} {\bibfnamefont {L.}~\bibnamefont {Berthier}}, \bibinfo
  {author} {\bibfnamefont {T.}~\bibnamefont {Divoux}}, \ and\ \bibinfo {author}
  {\bibfnamefont {S.}~\bibnamefont {Manneville}},\ }\href@noop {} {\bibfield
  {journal} {\bibinfo  {journal} {Reviews of Modern Physics}\ }\textbf
  {\bibinfo {volume} {89}},\ \bibinfo {pages} {035005} (\bibinfo {year}
  {2017})}\BibitemShut {NoStop}%
\bibitem [{\citenamefont {Pine}\ \emph {et~al.}(2005)\citenamefont {Pine},
  \citenamefont {Gollub}, \citenamefont {Brady},\ and\ \citenamefont
  {Leshansky}}]{Pine2005}%
  \BibitemOpen
  \bibfield  {author} {\bibinfo {author} {\bibfnamefont {D.~J.}\ \bibnamefont
  {Pine}}, \bibinfo {author} {\bibfnamefont {J.~P.}\ \bibnamefont {Gollub}},
  \bibinfo {author} {\bibfnamefont {J.~F.}\ \bibnamefont {Brady}}, \ and\
  \bibinfo {author} {\bibfnamefont {A.~M.}\ \bibnamefont {Leshansky}},\
  }\href@noop {} {\bibfield  {journal} {\bibinfo  {journal} {Nature}\ }\textbf
  {\bibinfo {volume} {438}},\ \bibinfo {pages} {997} (\bibinfo {year}
  {2005})}\BibitemShut {NoStop}%
\bibitem [{\citenamefont {Cort{\'e}}\ \emph {et~al.}(2008)\citenamefont
  {Cort{\'e}}, \citenamefont {Chaikin}, \citenamefont {Gollub},\ and\
  \citenamefont {Pine}}]{cortePineNature2008}%
  \BibitemOpen
  \bibfield  {author} {\bibinfo {author} {\bibfnamefont {L.}~\bibnamefont
  {Cort{\'e}}}, \bibinfo {author} {\bibfnamefont {P.~M.}\ \bibnamefont
  {Chaikin}}, \bibinfo {author} {\bibfnamefont {J.~P.}\ \bibnamefont {Gollub}},
  \ and\ \bibinfo {author} {\bibfnamefont {D.~J.}\ \bibnamefont {Pine}},\
  }\href@noop {} {\bibfield  {journal} {\bibinfo  {journal} {Nature Physics}\
  }\textbf {\bibinfo {volume} {4}},\ \bibinfo {pages} {420} (\bibinfo {year}
  {2008})}\BibitemShut {NoStop}%
\bibitem [{\citenamefont {Cort\'e}\ \emph {et~al.}(2009)\citenamefont
  {Cort\'e}, \citenamefont {Gerbode}, \citenamefont {Man},\ and\ \citenamefont
  {Pine}}]{CortePRL2009}%
  \BibitemOpen
  \bibfield  {author} {\bibinfo {author} {\bibfnamefont {L.}~\bibnamefont
  {Cort\'e}}, \bibinfo {author} {\bibfnamefont {S.~J.}\ \bibnamefont
  {Gerbode}}, \bibinfo {author} {\bibfnamefont {W.}~\bibnamefont {Man}}, \ and\
  \bibinfo {author} {\bibfnamefont {D.~J.}\ \bibnamefont {Pine}},\ }\href@noop
  {} {\bibfield  {journal} {\bibinfo  {journal} {Physical Review Letters}\
  }\textbf {\bibinfo {volume} {103}},\ \bibinfo {pages} {248301} (\bibinfo
  {year} {2009})}\BibitemShut {NoStop}%
\bibitem [{\citenamefont {Hima~Nagamanasa}\ \emph {et~al.}(2014)\citenamefont
  {Hima~Nagamanasa}, \citenamefont {Gokhale}, \citenamefont {Sood},\ and\
  \citenamefont {Ganapathy}}]{MansaKandulaPREA2014}%
  \BibitemOpen
  \bibfield  {author} {\bibinfo {author} {\bibfnamefont {K.}~\bibnamefont
  {Hima~Nagamanasa}}, \bibinfo {author} {\bibfnamefont {S.}~\bibnamefont
  {Gokhale}}, \bibinfo {author} {\bibfnamefont {A.~K.}\ \bibnamefont {Sood}}, \
  and\ \bibinfo {author} {\bibfnamefont {R.}~\bibnamefont {Ganapathy}},\
  }\href@noop {} {\bibfield  {journal} {\bibinfo  {journal} {Physical Review
  E}\ }\textbf {\bibinfo {volume} {89}},\ \bibinfo {pages} {062308} (\bibinfo
  {year} {2014})}\BibitemShut {NoStop}%
\bibitem [{\citenamefont {Fiocco}\ \emph {et~al.}(2013)\citenamefont {Fiocco},
  \citenamefont {Foffi},\ and\ \citenamefont
  {Sastry}}]{FioccoFoffiSastryPRE2013}%
  \BibitemOpen
  \bibfield  {author} {\bibinfo {author} {\bibfnamefont {D.}~\bibnamefont
  {Fiocco}}, \bibinfo {author} {\bibfnamefont {G.}~\bibnamefont {Foffi}}, \
  and\ \bibinfo {author} {\bibfnamefont {S.}~\bibnamefont {Sastry}},\
  }\href@noop {} {\bibfield  {journal} {\bibinfo  {journal} {Physical Review
  E}\ }\textbf {\bibinfo {volume} {88}},\ \bibinfo {pages} {020301} (\bibinfo
  {year} {2013})}\BibitemShut {NoStop}%
\bibitem [{\citenamefont {Bhaumik}\ \emph {et~al.}(2021)\citenamefont
  {Bhaumik}, \citenamefont {Foffi},\ and\ \citenamefont
  {Sastry}}]{HimangshuPNAS2021}%
  \BibitemOpen
  \bibfield  {author} {\bibinfo {author} {\bibfnamefont {H.}~\bibnamefont
  {Bhaumik}}, \bibinfo {author} {\bibfnamefont {G.}~\bibnamefont {Foffi}}, \
  and\ \bibinfo {author} {\bibfnamefont {S.}~\bibnamefont {Sastry}},\
  }\href@noop {} {\bibfield  {journal} {\bibinfo  {journal} {Proceedings of the
  National Academy of Sciences}\ }\textbf {\bibinfo {volume} {118}},\ \bibinfo
  {pages} {e2100227118} (\bibinfo {year} {2021})}\BibitemShut {NoStop}%
\bibitem [{\citenamefont {Bhowmik}\ \emph {et~al.}(2022)\citenamefont
  {Bhowmik}, \citenamefont {Hentchel},\ and\ \citenamefont
  {Procaccia}}]{BHPEPL2022}%
  \BibitemOpen
  \bibfield  {author} {\bibinfo {author} {\bibfnamefont {B.~P.}\ \bibnamefont
  {Bhowmik}}, \bibinfo {author} {\bibfnamefont {H.~G.~E.}\ \bibnamefont
  {Hentchel}}, \ and\ \bibinfo {author} {\bibfnamefont {I.}~\bibnamefont
  {Procaccia}},\ }\href@noop {} {\bibfield  {journal} {\bibinfo  {journal}
  {Europhysics Letters}\ }\textbf {\bibinfo {volume} {137}},\ \bibinfo {pages}
  {46002} (\bibinfo {year} {2022})}\BibitemShut {NoStop}%
\bibitem [{\citenamefont {Saitoh}\ and\ \citenamefont
  {Tighe}(2019)}]{TigheNonLocalPRL}%
  \BibitemOpen
  \bibfield  {author} {\bibinfo {author} {\bibfnamefont {K.}~\bibnamefont
  {Saitoh}}\ and\ \bibinfo {author} {\bibfnamefont {B.~P.}\ \bibnamefont
  {Tighe}},\ }\href@noop {} {\bibfield  {journal} {\bibinfo  {journal}
  {Physical Review Letters}\ }\textbf {\bibinfo {volume} {122}},\ \bibinfo
  {pages} {188001} (\bibinfo {year} {2019})}\BibitemShut {NoStop}%
\bibitem [{\citenamefont {Bhowmik}\ and\ \citenamefont
  {Ness}(2023)}]{bhowmik2023scaling}%
  \BibitemOpen
  \bibfield  {author} {\bibinfo {author} {\bibfnamefont {B.~P.}\ \bibnamefont
  {Bhowmik}}\ and\ \bibinfo {author} {\bibfnamefont {C.}~\bibnamefont {Ness}},\
  }\href@noop {} {\bibfield  {journal} {\bibinfo  {journal} {arXiv preprint
  arXiv:2308.08402}\ } (\bibinfo {year} {2023})}\BibitemShut {NoStop}%
\bibitem [{\citenamefont {Goyon}\ \emph {et~al.}(2008)\citenamefont {Goyon},
  \citenamefont {Colin}, \citenamefont {Ovarlez}, \citenamefont {Ajdari},\ and\
  \citenamefont {Bocquet}}]{BocquetNonLocalNature}%
  \BibitemOpen
  \bibfield  {author} {\bibinfo {author} {\bibfnamefont {J.}~\bibnamefont
  {Goyon}}, \bibinfo {author} {\bibfnamefont {A.}~\bibnamefont {Colin}},
  \bibinfo {author} {\bibfnamefont {G.}~\bibnamefont {Ovarlez}}, \bibinfo
  {author} {\bibfnamefont {A.}~\bibnamefont {Ajdari}}, \ and\ \bibinfo {author}
  {\bibfnamefont {L.}~\bibnamefont {Bocquet}},\ }\href@noop {} {\bibfield
  {journal} {\bibinfo  {journal} {Nature}\ }\textbf {\bibinfo {volume} {454}},\
  \bibinfo {pages} {84} (\bibinfo {year} {2008})}\BibitemShut {NoStop}%
\bibitem [{\citenamefont {Kamrin}\ and\ \citenamefont
  {Koval}(2012)}]{KamrinKovalPRL2012}%
  \BibitemOpen
  \bibfield  {author} {\bibinfo {author} {\bibfnamefont {K.}~\bibnamefont
  {Kamrin}}\ and\ \bibinfo {author} {\bibfnamefont {G.}~\bibnamefont {Koval}},\
  }\href@noop {} {\bibfield  {journal} {\bibinfo  {journal} {Physical Review
  Letters}\ }\textbf {\bibinfo {volume} {108}},\ \bibinfo {pages} {178301}
  (\bibinfo {year} {2012})}\BibitemShut {NoStop}%
\bibitem [{\citenamefont {Chaudhuri}\ and\ \citenamefont
  {Horbach}(2014)}]{PinakiAndJuergenPRE2014}%
  \BibitemOpen
  \bibfield  {author} {\bibinfo {author} {\bibfnamefont {P.}~\bibnamefont
  {Chaudhuri}}\ and\ \bibinfo {author} {\bibfnamefont {J.}~\bibnamefont
  {Horbach}},\ }\href@noop {} {\bibfield  {journal} {\bibinfo  {journal}
  {Physical Review E}\ }\textbf {\bibinfo {volume} {90}},\ \bibinfo {pages}
  {040301} (\bibinfo {year} {2014})}\BibitemShut {NoStop}%
\bibitem [{\citenamefont {Schreck}\ \emph {et~al.}(2013)\citenamefont
  {Schreck}, \citenamefont {Hoy}, \citenamefont {Shattuck},\ and\ \citenamefont
  {O’Hern}}]{schreck2013particle}%
  \BibitemOpen
  \bibfield  {author} {\bibinfo {author} {\bibfnamefont {C.~F.}\ \bibnamefont
  {Schreck}}, \bibinfo {author} {\bibfnamefont {R.~S.}\ \bibnamefont {Hoy}},
  \bibinfo {author} {\bibfnamefont {M.~D.}\ \bibnamefont {Shattuck}}, \ and\
  \bibinfo {author} {\bibfnamefont {C.~S.}\ \bibnamefont {O’Hern}},\
  }\href@noop {} {\bibfield  {journal} {\bibinfo  {journal} {Physical Review
  E}\ }\textbf {\bibinfo {volume} {88}},\ \bibinfo {pages} {052205} (\bibinfo
  {year} {2013})}\BibitemShut {NoStop}%
\bibitem [{\citenamefont {Tjhung}\ and\ \citenamefont
  {Berthier}(2015)}]{TjhungAndBethierPRL2015}%
  \BibitemOpen
  \bibfield  {author} {\bibinfo {author} {\bibfnamefont {E.}~\bibnamefont
  {Tjhung}}\ and\ \bibinfo {author} {\bibfnamefont {L.}~\bibnamefont
  {Berthier}},\ }\href@noop {} {\bibfield  {journal} {\bibinfo  {journal}
  {Physical Review Letters}\ }\textbf {\bibinfo {volume} {114}},\ \bibinfo
  {pages} {148301} (\bibinfo {year} {2015})}\BibitemShut {NoStop}%
\bibitem [{\citenamefont {Galliano}\ \emph {et~al.}(2023)\citenamefont
  {Galliano}, \citenamefont {Cates},\ and\ \citenamefont
  {Berthier}}]{GCBPRL2023}%
  \BibitemOpen
  \bibfield  {author} {\bibinfo {author} {\bibfnamefont {L.}~\bibnamefont
  {Galliano}}, \bibinfo {author} {\bibfnamefont {M.~E.}\ \bibnamefont {Cates}},
  \ and\ \bibinfo {author} {\bibfnamefont {L.}~\bibnamefont {Berthier}},\
  }\href@noop {} {\bibfield  {journal} {\bibinfo  {journal} {Physical Review
  Letters}\ }\textbf {\bibinfo {volume} {131}},\ \bibinfo {pages} {047101}
  (\bibinfo {year} {2023})}\BibitemShut {NoStop}%
\bibitem [{\citenamefont {Hexner}\ and\ \citenamefont
  {Levine}(2015)}]{HexnerPRL2015}%
  \BibitemOpen
  \bibfield  {author} {\bibinfo {author} {\bibfnamefont {D.}~\bibnamefont
  {Hexner}}\ and\ \bibinfo {author} {\bibfnamefont {D.}~\bibnamefont
  {Levine}},\ }\href@noop {} {\bibfield  {journal} {\bibinfo  {journal}
  {Physical Review Letters}\ }\textbf {\bibinfo {volume} {114}},\ \bibinfo
  {pages} {110602} (\bibinfo {year} {2015})}\BibitemShut {NoStop}%
\bibitem [{\citenamefont {Ness}\ and\ \citenamefont
  {Cates}(2020)}]{ChrisNessPRL2020AbsState}%
  \BibitemOpen
  \bibfield  {author} {\bibinfo {author} {\bibfnamefont {C.}~\bibnamefont
  {Ness}}\ and\ \bibinfo {author} {\bibfnamefont {M.~E.}\ \bibnamefont
  {Cates}},\ }\href@noop {} {\bibfield  {journal} {\bibinfo  {journal}
  {Physical Review Letters}\ }\textbf {\bibinfo {volume} {124}},\ \bibinfo
  {pages} {088004} (\bibinfo {year} {2020})}\BibitemShut {NoStop}%
\bibitem [{\citenamefont {Ghosh}\ \emph {et~al.}(2022)\citenamefont {Ghosh},
  \citenamefont {Radhakrishnan}, \citenamefont {Chaikin}, \citenamefont
  {Levine},\ and\ \citenamefont {Ghosh}}]{ShankarGhoshPRL2022}%
  \BibitemOpen
  \bibfield  {author} {\bibinfo {author} {\bibfnamefont {A.}~\bibnamefont
  {Ghosh}}, \bibinfo {author} {\bibfnamefont {J.}~\bibnamefont
  {Radhakrishnan}}, \bibinfo {author} {\bibfnamefont {P.~M.}\ \bibnamefont
  {Chaikin}}, \bibinfo {author} {\bibfnamefont {D.}~\bibnamefont {Levine}}, \
  and\ \bibinfo {author} {\bibfnamefont {S.}~\bibnamefont {Ghosh}},\
  }\href@noop {} {\bibfield  {journal} {\bibinfo  {journal} {Physical Review
  Letters}\ }\textbf {\bibinfo {volume} {129}},\ \bibinfo {pages} {188002}
  (\bibinfo {year} {2022})}\BibitemShut {NoStop}%
\bibitem [{\citenamefont {Mari}\ \emph {et~al.}(2022)\citenamefont {Mari},
  \citenamefont {Bertin},\ and\ \citenamefont {Nardini}}]{MariPRE2022}%
  \BibitemOpen
  \bibfield  {author} {\bibinfo {author} {\bibfnamefont {R.}~\bibnamefont
  {Mari}}, \bibinfo {author} {\bibfnamefont {E.}~\bibnamefont {Bertin}}, \ and\
  \bibinfo {author} {\bibfnamefont {C.}~\bibnamefont {Nardini}},\ }\href@noop
  {} {\bibfield  {journal} {\bibinfo  {journal} {Physical Review E}\ }\textbf
  {\bibinfo {volume} {105}},\ \bibinfo {pages} {L032602} (\bibinfo {year}
  {2022})}\BibitemShut {NoStop}%
\bibitem [{\citenamefont {Milz}\ and\ \citenamefont
  {Schmiedeberg}(2013)}]{MilzSchmiedebergPRE2013}%
  \BibitemOpen
  \bibfield  {author} {\bibinfo {author} {\bibfnamefont {L.}~\bibnamefont
  {Milz}}\ and\ \bibinfo {author} {\bibfnamefont {M.}~\bibnamefont
  {Schmiedeberg}},\ }\href@noop {} {\bibfield  {journal} {\bibinfo  {journal}
  {Physical Review E}\ }\textbf {\bibinfo {volume} {88}},\ \bibinfo {pages}
  {062308} (\bibinfo {year} {2013})}\BibitemShut {NoStop}%
\bibitem [{\citenamefont {L{\"u}beck}(2004)}]{SLuebeckScaling2004}%
  \BibitemOpen
  \bibfield  {author} {\bibinfo {author} {\bibfnamefont {S.}~\bibnamefont
  {L{\"u}beck}},\ }\href@noop {} {\bibfield  {journal} {\bibinfo  {journal}
  {International Journal of Modern Physics B}\ }\textbf {\bibinfo {volume}
  {18}},\ \bibinfo {pages} {3977} (\bibinfo {year} {2004})}\BibitemShut
  {NoStop}%
\bibitem [{\citenamefont {Menon}\ and\ \citenamefont
  {Ramaswamy}(2009)}]{SriramGautomMenonPRE}%
  \BibitemOpen
  \bibfield  {author} {\bibinfo {author} {\bibfnamefont {G.~I.}\ \bibnamefont
  {Menon}}\ and\ \bibinfo {author} {\bibfnamefont {S.}~\bibnamefont
  {Ramaswamy}},\ }\href@noop {} {\bibfield  {journal} {\bibinfo  {journal}
  {Physical Review E}\ }\textbf {\bibinfo {volume} {79}},\ \bibinfo {pages}
  {061108} (\bibinfo {year} {2009})}\BibitemShut {NoStop}%
\bibitem [{\citenamefont {Pham}\ \emph {et~al.}(2016)\citenamefont {Pham},
  \citenamefont {Butler},\ and\ \citenamefont {Metzger}}]{pham2016origin}%
  \BibitemOpen
  \bibfield  {author} {\bibinfo {author} {\bibfnamefont {P.}~\bibnamefont
  {Pham}}, \bibinfo {author} {\bibfnamefont {J.~E.}\ \bibnamefont {Butler}}, \
  and\ \bibinfo {author} {\bibfnamefont {B.}~\bibnamefont {Metzger}},\
  }\href@noop {} {\bibfield  {journal} {\bibinfo  {journal} {Physical Review
  Fluids}\ }\textbf {\bibinfo {volume} {1}},\ \bibinfo {pages} {022201}
  (\bibinfo {year} {2016})}\BibitemShut {NoStop}%
\bibitem [{\citenamefont {Oh}\ \emph {et~al.}(2015)\citenamefont {Oh},
  \citenamefont {Song}, \citenamefont {Garagash}, \citenamefont {Lecampion},\
  and\ \citenamefont {Desroches}}]{Migration2}%
  \BibitemOpen
  \bibfield  {author} {\bibinfo {author} {\bibfnamefont {S.}~\bibnamefont
  {Oh}}, \bibinfo {author} {\bibfnamefont {Y.-q.}\ \bibnamefont {Song}},
  \bibinfo {author} {\bibfnamefont {D.~I.}\ \bibnamefont {Garagash}}, \bibinfo
  {author} {\bibfnamefont {B.}~\bibnamefont {Lecampion}}, \ and\ \bibinfo
  {author} {\bibfnamefont {J.}~\bibnamefont {Desroches}},\ }\href@noop {}
  {\bibfield  {journal} {\bibinfo  {journal} {Physical Review Letters}\
  }\textbf {\bibinfo {volume} {114}},\ \bibinfo {pages} {088301} (\bibinfo
  {year} {2015})}\BibitemShut {NoStop}%
\bibitem [{\citenamefont {Hampton}\ \emph {et~al.}(1997)\citenamefont
  {Hampton}, \citenamefont {Mammoli}, \citenamefont {Graham}, \citenamefont
  {Tetlow},\ and\ \citenamefont {Altobelli}}]{Migration1}%
  \BibitemOpen
  \bibfield  {author} {\bibinfo {author} {\bibfnamefont {R.~E.}\ \bibnamefont
  {Hampton}}, \bibinfo {author} {\bibfnamefont {A.~A.}\ \bibnamefont
  {Mammoli}}, \bibinfo {author} {\bibfnamefont {A.~L.}\ \bibnamefont {Graham}},
  \bibinfo {author} {\bibfnamefont {N.}~\bibnamefont {Tetlow}}, \ and\ \bibinfo
  {author} {\bibfnamefont {S.~A.}\ \bibnamefont {Altobelli}},\ }\href@noop {}
  {\bibfield  {journal} {\bibinfo  {journal} {Journal of Rheology}\ }\textbf
  {\bibinfo {volume} {41}},\ \bibinfo {pages} {621} (\bibinfo {year}
  {1997})}\BibitemShut {NoStop}%
\bibitem [{\citenamefont {Gillissen}\ and\ \citenamefont
  {Ness}(2020)}]{GillissenNessPRL2020}%
  \BibitemOpen
  \bibfield  {author} {\bibinfo {author} {\bibfnamefont {J.~J.~J.}\
  \bibnamefont {Gillissen}}\ and\ \bibinfo {author} {\bibfnamefont
  {C.}~\bibnamefont {Ness}},\ }\href@noop {} {\bibfield  {journal} {\bibinfo
  {journal} {Physical Review Letters}\ }\textbf {\bibinfo {volume} {125}},\
  \bibinfo {pages} {184503} (\bibinfo {year} {2020})}\BibitemShut {NoStop}%
\bibitem [{\citenamefont {Vespignani}\ and\ \citenamefont
  {Zapperi}(1997)}]{VespignaniZapperiPRL1997}%
  \BibitemOpen
  \bibfield  {author} {\bibinfo {author} {\bibfnamefont {A.}~\bibnamefont
  {Vespignani}}\ and\ \bibinfo {author} {\bibfnamefont {S.}~\bibnamefont
  {Zapperi}},\ }\href@noop {} {\bibfield  {journal} {\bibinfo  {journal}
  {Physical Review Letters}\ }\textbf {\bibinfo {volume} {78}},\ \bibinfo
  {pages} {4793} (\bibinfo {year} {1997})}\BibitemShut {NoStop}%
\bibitem [{\citenamefont {Vespignani}\ \emph {et~al.}(1998)\citenamefont
  {Vespignani}, \citenamefont {Dickman}, \citenamefont {Mu\~noz},\ and\
  \citenamefont {Zapperi}}]{ZapperiPRL1998}%
  \BibitemOpen
  \bibfield  {author} {\bibinfo {author} {\bibfnamefont {A.}~\bibnamefont
  {Vespignani}}, \bibinfo {author} {\bibfnamefont {R.}~\bibnamefont {Dickman}},
  \bibinfo {author} {\bibfnamefont {M.~A.}\ \bibnamefont {Mu\~noz}}, \ and\
  \bibinfo {author} {\bibfnamefont {S.}~\bibnamefont {Zapperi}},\ }\href@noop
  {} {\bibfield  {journal} {\bibinfo  {journal} {Physical Review Letters}\
  }\textbf {\bibinfo {volume} {81}},\ \bibinfo {pages} {5676} (\bibinfo {year}
  {1998})}\BibitemShut {NoStop}%
\bibitem [{\citenamefont {Hipke}\ \emph {et~al.}(2009)\citenamefont {Hipke},
  \citenamefont {Lübeck},\ and\ \citenamefont {Hinrichsen}}]{Hipke2009}%
  \BibitemOpen
  \bibfield  {author} {\bibinfo {author} {\bibfnamefont {A.}~\bibnamefont
  {Hipke}}, \bibinfo {author} {\bibfnamefont {S.}~\bibnamefont {Lübeck}}, \
  and\ \bibinfo {author} {\bibfnamefont {H.}~\bibnamefont {Hinrichsen}},\
  }\href@noop {} {\bibfield  {journal} {\bibinfo  {journal} {Journal of
  Statistical Mechanics: Theory and Experiment}\ }\textbf {\bibinfo {volume}
  {2009}},\ \bibinfo {pages} {P07021} (\bibinfo {year} {2009})}\BibitemShut
  {NoStop}%
\bibitem [{\citenamefont {Xu}\ and\ \citenamefont
  {Schwarz}(2013)}]{XuSchwarzPRE2013}%
  \BibitemOpen
  \bibfield  {author} {\bibinfo {author} {\bibfnamefont {S.-L.-Y.}\
  \bibnamefont {Xu}}\ and\ \bibinfo {author} {\bibfnamefont {J.~M.}\
  \bibnamefont {Schwarz}},\ }\href@noop {} {\bibfield  {journal} {\bibinfo
  {journal} {Physical Review E}\ }\textbf {\bibinfo {volume} {88}},\ \bibinfo
  {pages} {052130} (\bibinfo {year} {2013})}\BibitemShut {NoStop}%
\bibitem [{\citenamefont {Cammarota}\ and\ \citenamefont
  {Biroli}(2012)}]{BiroliCammarotaPNAS2005}%
  \BibitemOpen
  \bibfield  {author} {\bibinfo {author} {\bibfnamefont {C.}~\bibnamefont
  {Cammarota}}\ and\ \bibinfo {author} {\bibfnamefont {G.}~\bibnamefont
  {Biroli}},\ }\href@noop {} {\bibfield  {journal} {\bibinfo  {journal}
  {Proceedings of the National Academy of Sciences}\ }\textbf {\bibinfo
  {volume} {109}},\ \bibinfo {pages} {8850} (\bibinfo {year}
  {2012})}\BibitemShut {NoStop}%
\bibitem [{\citenamefont {Das}\ \emph {et~al.}(2023)\citenamefont {Das},
  \citenamefont {Bhowmik}, \citenamefont {Puthirath}, \citenamefont
  {Narayanan},\ and\ \citenamefont {Karmakar}}]{DBPNKPNASNexus2023}%
  \BibitemOpen
  \bibfield  {author} {\bibinfo {author} {\bibfnamefont {R.}~\bibnamefont
  {Das}}, \bibinfo {author} {\bibfnamefont {B.~P.}\ \bibnamefont {Bhowmik}},
  \bibinfo {author} {\bibfnamefont {A.~B.}\ \bibnamefont {Puthirath}}, \bibinfo
  {author} {\bibfnamefont {T.~N.}\ \bibnamefont {Narayanan}}, \ and\ \bibinfo
  {author} {\bibfnamefont {S.}~\bibnamefont {Karmakar}},\ }\href@noop {}
  {\bibfield  {journal} {\bibinfo  {journal} {PNAS Nexus}\ }\textbf {\bibinfo
  {volume} {2}},\ \bibinfo {pages} {pgad277} (\bibinfo {year}
  {2023})}\BibitemShut {NoStop}%
\bibitem [{\citenamefont {Bhowmik}\ \emph
  {et~al.}(2019{\natexlab{a}})\citenamefont {Bhowmik}, \citenamefont
  {Chaudhuri},\ and\ \citenamefont {Karmakar}}]{BCKPRL2019}%
  \BibitemOpen
  \bibfield  {author} {\bibinfo {author} {\bibfnamefont {B.~P.}\ \bibnamefont
  {Bhowmik}}, \bibinfo {author} {\bibfnamefont {P.}~\bibnamefont {Chaudhuri}},
  \ and\ \bibinfo {author} {\bibfnamefont {S.}~\bibnamefont {Karmakar}},\
  }\href@noop {} {\bibfield  {journal} {\bibinfo  {journal} {Physical Review
  Letters}\ }\textbf {\bibinfo {volume} {123}},\ \bibinfo {pages} {185501}
  (\bibinfo {year} {2019}{\natexlab{a}})}\BibitemShut {NoStop}%
\bibitem [{\citenamefont {Bhowmik}\ \emph
  {et~al.}(2019{\natexlab{b}})\citenamefont {Bhowmik}, \citenamefont
  {Karmakar}, \citenamefont {Procaccia},\ and\ \citenamefont
  {Rainone}}]{BKPRPRE2019}%
  \BibitemOpen
  \bibfield  {author} {\bibinfo {author} {\bibfnamefont {B.~P.}\ \bibnamefont
  {Bhowmik}}, \bibinfo {author} {\bibfnamefont {S.}~\bibnamefont {Karmakar}},
  \bibinfo {author} {\bibfnamefont {I.}~\bibnamefont {Procaccia}}, \ and\
  \bibinfo {author} {\bibfnamefont {C.}~\bibnamefont {Rainone}},\ }\href@noop
  {} {\bibfield  {journal} {\bibinfo  {journal} {Physical Review E}\ }\textbf
  {\bibinfo {volume} {100}},\ \bibinfo {pages} {052110} (\bibinfo {year}
  {2019}{\natexlab{b}})}\BibitemShut {NoStop}%
\end{thebibliography}%
\end{document}